\documentclass[letterpaper,compsoc,twoside]{IEEEtran}
\usepackage{fixltx2e} 
\usepackage{cmap} 
\usepackage{ifthen}
\usepackage[T1]{fontenc}
\usepackage[utf8]{inputenc}
\usepackage{amsmath}

\usepackage[font={small,it},labelfont=bf]{caption}
\usepackage{float}

\pdfoutput=1
\usepackage{scipy}
\makeatletter
\def\PY@reset{\let\PY@it=\relax \let\PY@bf=\relax%
    \let\PY@ul=\relax \let\PY@tc=\relax%
    \let\PY@bc=\relax \let\PY@ff=\relax}
\def\PY@tok#1{\csname PY@tok@#1\endcsname}
\def\PY@toks#1+{\ifx\relax#1\empty\else%
    \PY@tok{#1}\expandafter\PY@toks\fi}
\def\PY@do#1{\PY@bc{\PY@tc{\PY@ul{%
    \PY@it{\PY@bf{\PY@ff{#1}}}}}}}
\def\PY#1#2{\PY@reset\PY@toks#1+\relax+\PY@do{#2}}

\expandafter\def\csname PY@tok@gd\endcsname{\def\PY@tc##1{\textcolor[rgb]{0.63,0.00,0.00}{##1}}}
\expandafter\def\csname PY@tok@gu\endcsname{\let\PY@bf=\textbf\def\PY@tc##1{\textcolor[rgb]{0.50,0.00,0.50}{##1}}}
\expandafter\def\csname PY@tok@gt\endcsname{\def\PY@tc##1{\textcolor[rgb]{0.00,0.27,0.87}{##1}}}
\expandafter\def\csname PY@tok@gs\endcsname{\let\PY@bf=\textbf}
\expandafter\def\csname PY@tok@gr\endcsname{\def\PY@tc##1{\textcolor[rgb]{1.00,0.00,0.00}{##1}}}
\expandafter\def\csname PY@tok@cm\endcsname{\let\PY@it=\textit\def\PY@tc##1{\textcolor[rgb]{0.25,0.50,0.56}{##1}}}
\expandafter\def\csname PY@tok@vg\endcsname{\def\PY@tc##1{\textcolor[rgb]{0.73,0.38,0.84}{##1}}}
\expandafter\def\csname PY@tok@m\endcsname{\def\PY@tc##1{\textcolor[rgb]{0.13,0.50,0.31}{##1}}}
\expandafter\def\csname PY@tok@mh\endcsname{\def\PY@tc##1{\textcolor[rgb]{0.13,0.50,0.31}{##1}}}
\expandafter\def\csname PY@tok@cs\endcsname{\def\PY@tc##1{\textcolor[rgb]{0.25,0.50,0.56}{##1}}\def\PY@bc##1{\setlength{\fboxsep}{0pt}\colorbox[rgb]{1.00,0.94,0.94}{\strut ##1}}}
\expandafter\def\csname PY@tok@ge\endcsname{\let\PY@it=\textit}
\expandafter\def\csname PY@tok@vc\endcsname{\def\PY@tc##1{\textcolor[rgb]{0.73,0.38,0.84}{##1}}}
\expandafter\def\csname PY@tok@il\endcsname{\def\PY@tc##1{\textcolor[rgb]{0.13,0.50,0.31}{##1}}}
\expandafter\def\csname PY@tok@go\endcsname{\def\PY@tc##1{\textcolor[rgb]{0.20,0.20,0.20}{##1}}}
\expandafter\def\csname PY@tok@cp\endcsname{\def\PY@tc##1{\textcolor[rgb]{0.00,0.44,0.13}{##1}}}
\expandafter\def\csname PY@tok@gi\endcsname{\def\PY@tc##1{\textcolor[rgb]{0.00,0.63,0.00}{##1}}}
\expandafter\def\csname PY@tok@gh\endcsname{\let\PY@bf=\textbf\def\PY@tc##1{\textcolor[rgb]{0.00,0.00,0.50}{##1}}}
\expandafter\def\csname PY@tok@ni\endcsname{\let\PY@bf=\textbf\def\PY@tc##1{\textcolor[rgb]{0.84,0.33,0.22}{##1}}}
\expandafter\def\csname PY@tok@nl\endcsname{\let\PY@bf=\textbf\def\PY@tc##1{\textcolor[rgb]{0.00,0.13,0.44}{##1}}}
\expandafter\def\csname PY@tok@nn\endcsname{\let\PY@bf=\textbf\def\PY@tc##1{\textcolor[rgb]{0.05,0.52,0.71}{##1}}}
\expandafter\def\csname PY@tok@no\endcsname{\def\PY@tc##1{\textcolor[rgb]{0.38,0.68,0.84}{##1}}}
\expandafter\def\csname PY@tok@na\endcsname{\def\PY@tc##1{\textcolor[rgb]{0.25,0.44,0.63}{##1}}}
\expandafter\def\csname PY@tok@nb\endcsname{\def\PY@tc##1{\textcolor[rgb]{0.00,0.44,0.13}{##1}}}
\expandafter\def\csname PY@tok@nc\endcsname{\let\PY@bf=\textbf\def\PY@tc##1{\textcolor[rgb]{0.05,0.52,0.71}{##1}}}
\expandafter\def\csname PY@tok@nd\endcsname{\let\PY@bf=\textbf\def\PY@tc##1{\textcolor[rgb]{0.33,0.33,0.33}{##1}}}
\expandafter\def\csname PY@tok@ne\endcsname{\def\PY@tc##1{\textcolor[rgb]{0.00,0.44,0.13}{##1}}}
\expandafter\def\csname PY@tok@nf\endcsname{\def\PY@tc##1{\textcolor[rgb]{0.02,0.16,0.49}{##1}}}
\expandafter\def\csname PY@tok@si\endcsname{\let\PY@it=\textit\def\PY@tc##1{\textcolor[rgb]{0.44,0.63,0.82}{##1}}}
\expandafter\def\csname PY@tok@s2\endcsname{\def\PY@tc##1{\textcolor[rgb]{0.25,0.44,0.63}{##1}}}
\expandafter\def\csname PY@tok@vi\endcsname{\def\PY@tc##1{\textcolor[rgb]{0.73,0.38,0.84}{##1}}}
\expandafter\def\csname PY@tok@nt\endcsname{\let\PY@bf=\textbf\def\PY@tc##1{\textcolor[rgb]{0.02,0.16,0.45}{##1}}}
\expandafter\def\csname PY@tok@nv\endcsname{\def\PY@tc##1{\textcolor[rgb]{0.73,0.38,0.84}{##1}}}
\expandafter\def\csname PY@tok@s1\endcsname{\def\PY@tc##1{\textcolor[rgb]{0.25,0.44,0.63}{##1}}}
\expandafter\def\csname PY@tok@gp\endcsname{\let\PY@bf=\textbf\def\PY@tc##1{\textcolor[rgb]{0.78,0.36,0.04}{##1}}}
\expandafter\def\csname PY@tok@sh\endcsname{\def\PY@tc##1{\textcolor[rgb]{0.25,0.44,0.63}{##1}}}
\expandafter\def\csname PY@tok@ow\endcsname{\let\PY@bf=\textbf\def\PY@tc##1{\textcolor[rgb]{0.00,0.44,0.13}{##1}}}
\expandafter\def\csname PY@tok@sx\endcsname{\def\PY@tc##1{\textcolor[rgb]{0.78,0.36,0.04}{##1}}}
\expandafter\def\csname PY@tok@bp\endcsname{\def\PY@tc##1{\textcolor[rgb]{0.00,0.44,0.13}{##1}}}
\expandafter\def\csname PY@tok@c1\endcsname{\let\PY@it=\textit\def\PY@tc##1{\textcolor[rgb]{0.25,0.50,0.56}{##1}}}
\expandafter\def\csname PY@tok@kc\endcsname{\let\PY@bf=\textbf\def\PY@tc##1{\textcolor[rgb]{0.00,0.44,0.13}{##1}}}
\expandafter\def\csname PY@tok@c\endcsname{\let\PY@it=\textit\def\PY@tc##1{\textcolor[rgb]{0.25,0.50,0.56}{##1}}}
\expandafter\def\csname PY@tok@mf\endcsname{\def\PY@tc##1{\textcolor[rgb]{0.13,0.50,0.31}{##1}}}
\expandafter\def\csname PY@tok@err\endcsname{\def\PY@bc##1{\setlength{\fboxsep}{0pt}\fcolorbox[rgb]{1.00,0.00,0.00}{1,1,1}{\strut ##1}}}
\expandafter\def\csname PY@tok@kd\endcsname{\let\PY@bf=\textbf\def\PY@tc##1{\textcolor[rgb]{0.00,0.44,0.13}{##1}}}
\expandafter\def\csname PY@tok@ss\endcsname{\def\PY@tc##1{\textcolor[rgb]{0.32,0.47,0.09}{##1}}}
\expandafter\def\csname PY@tok@sr\endcsname{\def\PY@tc##1{\textcolor[rgb]{0.14,0.33,0.53}{##1}}}
\expandafter\def\csname PY@tok@mo\endcsname{\def\PY@tc##1{\textcolor[rgb]{0.13,0.50,0.31}{##1}}}
\expandafter\def\csname PY@tok@mi\endcsname{\def\PY@tc##1{\textcolor[rgb]{0.13,0.50,0.31}{##1}}}
\expandafter\def\csname PY@tok@kn\endcsname{\let\PY@bf=\textbf\def\PY@tc##1{\textcolor[rgb]{0.00,0.44,0.13}{##1}}}
\expandafter\def\csname PY@tok@o\endcsname{\def\PY@tc##1{\textcolor[rgb]{0.40,0.40,0.40}{##1}}}
\expandafter\def\csname PY@tok@kr\endcsname{\let\PY@bf=\textbf\def\PY@tc##1{\textcolor[rgb]{0.00,0.44,0.13}{##1}}}
\expandafter\def\csname PY@tok@s\endcsname{\def\PY@tc##1{\textcolor[rgb]{0.25,0.44,0.63}{##1}}}
\expandafter\def\csname PY@tok@kp\endcsname{\def\PY@tc##1{\textcolor[rgb]{0.00,0.44,0.13}{##1}}}
\expandafter\def\csname PY@tok@w\endcsname{\def\PY@tc##1{\textcolor[rgb]{0.73,0.73,0.73}{##1}}}
\expandafter\def\csname PY@tok@kt\endcsname{\def\PY@tc##1{\textcolor[rgb]{0.56,0.13,0.00}{##1}}}
\expandafter\def\csname PY@tok@sc\endcsname{\def\PY@tc##1{\textcolor[rgb]{0.25,0.44,0.63}{##1}}}
\expandafter\def\csname PY@tok@sb\endcsname{\def\PY@tc##1{\textcolor[rgb]{0.25,0.44,0.63}{##1}}}
\expandafter\def\csname PY@tok@k\endcsname{\let\PY@bf=\textbf\def\PY@tc##1{\textcolor[rgb]{0.00,0.44,0.13}{##1}}}
\expandafter\def\csname PY@tok@se\endcsname{\let\PY@bf=\textbf\def\PY@tc##1{\textcolor[rgb]{0.25,0.44,0.63}{##1}}}
\expandafter\def\csname PY@tok@sd\endcsname{\let\PY@it=\textit\def\PY@tc##1{\textcolor[rgb]{0.25,0.44,0.63}{##1}}}

\def\PYZus{\char`\_}
\def\PYZob{\char`\{}
\def\PYZcb{\char`\}}

\def\PYZlt{\char`\<}
\def\PYZgt{\char`\>}
\def\PYZsh{\char`\#}

\def\PYZhy{\char`\-}
\def\PYZsq{\char`\'}

\makeatother

\providecommand*{\DUrole}[2]{%
  \ifcsname DUrole#1\endcsname%
    \csname DUrole#1\endcsname{#2}%
  \else
    \ifcsname docutilsrole#1\endcsname%
      \csname docutilsrole#1\endcsname{#2}%
    \else%
      #2%
    \fi%
  \fi%
}

\ifthenelse{\isundefined{\hypersetup}}{
  \usepackage[colorlinks=true,linkcolor=blue,urlcolor=blue]{hyperref}
  \urlstyle{same} 
}{}

\begin{document}
\newcounter{footnotecounter}\title{SfePy - Write Your Own FE Application}\author{Robert Cimrman$^{\setcounter{footnotecounter}{1}\fnsymbol{footnotecounter}\setcounter{footnotecounter}{2}\fnsymbol{footnotecounter}}$%
          \setcounter{footnotecounter}{1}\thanks{\fnsymbol{footnotecounter} %
          Corresponding author: \protect\href{mailto:cimrman3@ntc.zcu.cz}{cimrman3@ntc.zcu.cz}}\setcounter{footnotecounter}{2}\thanks{\fnsymbol{footnotecounter} New Technologies Research Centre, University of West Bohemia,
Plzeň, Czech Republic}\thanks{%

          \noindent%
          Copyright\,\copyright\,2014 Robert Cimrman. This is an open-access article distributed under the terms of the Creative Commons Attribution License, which permits unrestricted use, distribution, and reproduction in any medium, provided the original author and source are credited. http://creativecommons.org/licenses/by/3.0/%
        }}\maketitle
          \renewcommand{\leftmark}{PROC. OF THE 6th EUR. CONF. ON PYTHON IN SCIENCE (EUROSCIPY 2013)}
          \renewcommand{\rightmark}{SFEPY - WRITE YOUR OWN FE APPLICATION}

\setcounter{page}{65}
\newcommand*{\docutilsroleref}{\ref}
\newcommand*{\docutilsrolelabel}{\label}
\AtEndDocument{\cleardoublepage}
\begin{abstract}SfePy (Simple Finite Elements in Python) is a framework for solving various
kinds of problems (mechanics, physics, biology, ...) described by partial
differential equations in two or three space dimensions by the finite
element method. The paper illustrates its use in an interactive environment
or as a framework for building custom finite-element based solvers.\end{abstract}\begin{IEEEkeywords}partial differential equations, finite element method, Python\end{IEEEkeywords}

\section{Introduction%
  \label{introduction}%
}

SfePy (Simple Finite Elements in Python) is a multi-platform (Linux, Mac OS X,
Windows) software released under the New BSD license, see \url{http://sfepy.org}. It
implements one of the standard ways of discretizing partial differential
equations (PDEs) which is the finite element method (FEM) \cite{R1}.

So far, SfePy has been employed for modelling in materials science, including,
for example, multiscale biomechanical modelling (bone, muscle tissue with blood
perfusion) \cite{R2}, \cite{R3}, \cite{R4}, \cite{R5}, \cite{R6}, \cite{R7}, computation of acoustic
transmission coefficients across an interface of arbitrary microstructure
geometry \cite{R8}, computation of phononic band gaps \cite{R9}, \cite{R10}, finite element
formulation of Schroedinger equation \cite{R11}, and other applications, see also
Figure \DUrole{ref}{gallery}. The software can be used as%
\begin{itemize}

\item 

a collection of modules (a library) for building custom or domain-specific
applications,
\item 

a highly configurable \textquotedbl{}black box\textquotedbl{} PDE solver with problem description files
in Python.
\end{itemize}

In this paper we focus on illustrating the former use by using a particular
example. All examples presented below were tested to work with the version
2013.3 of SfePy.

\section{Development%
  \label{development}%
}

The SfePy project uses Git for source code management and GitHub web site for
the source code hosting and developer interaction, similarly to many other
scientific python tools. The version 2013.3 has been released in
September 18. 2013, and its git-controlled sources contained 761 total files,
138191 total lines (725935 added, 587744 removed) and 3857 commits done by 15
authors. However, about 90\% of all commits were done by the main developer as
most authors contributed only a few commits. The project seeks and encourages
more regular contributors, see \url{http://sfepy.org/doc-devel/development.html}.

\section{Short Description%
  \label{short-description}%
}

The code is written mostly in Python. For speed in general, it relies on fast
vectorized operations provided by NumPy \cite{R12} arrays, with heavy use of
advanced broadcasting and \textquotedbl{}index tricks\textquotedbl{} features. C and Cython \cite{R13} are used
in places where vectorization is not possible, or is too
difficult/unreadable. Other components of the scientific Python software stack
are used as well, among others: SciPy \cite{R14} solvers and algorithms, Matplotlib
\cite{R15} for 2D plots, Mayavi \cite{R16} for 3D plots and simple postprocessing GUI,
IPython \cite{R17} for a customized shell, SymPy \cite{R18} for symbolic
operations/code generation etc.

The basic structure of the code allows a flexible definition of various
problems. The problems are defined using components directly corresponding to
mathematical counterparts present in a weak formulation of a problem in the
finite element setting: a solution domain and its sub-domains (regions),
variables belonging to suitable discrete function spaces, equations as sums
of terms (weak form integrals), various kinds of boundary conditions,
material/constitutive parameters etc.

The key notion in SfePy is a \emph{term}, which is the smallest unit that can be
used to build \emph{equations} (linear combinations of terms). It corresponds to a
weak formulation integral and takes usually several arguments: (optional)
material parameters, a single virtual (or test) function variable and zero or
more state (or unknown) variables. The available terms (currently 105) are
listed at our web site (\url{http://sfepy.org/doc-devel/terms_overview.html}). The
already existing terms allow to solve problems from many scientific domains,
see Figure \DUrole{ref}{gallery}. Those terms cover many common PDEs in continuum
mechanics, poromechanics, biomechanics etc. with a notable exception of
electromagnetism (work in progress, see below).

SfePy discretizes PDEs using a continuous Galerkin approximation with finite
elements as defined in \cite{R1}. Discontinuous element-wise constant approximation
is also possible. Currently the code supports the 2D area (triangle, rectangle)
and 3D volume (tetrahedron, hexahedron) elements. Structural elements like
shells, plates, membranes or beams are not supported with a single exception of
a hyperelastic Mooney-Rivlin membrane.

Several kinds of basis or shape functions can be used for the finite element
approximation of the physical fields:%
\begin{itemize}

\item 

the classical nodal (Lagrange) basis can be used with all element types;
\item 

the hierarchical (Lobatto) basis can be used with tensor-product elements
(rectangle, hexahedron).
\end{itemize}

Although the code can provide the basis function polynomials of a high order
(10 or more), orders greater than 4 are not practically usable, especially in
3D. This is caused by the way the code assembles the element contributions into
the global matrix - it relies on NumPy vectorization and evaluates the element
matrices all at once and then adds to the global matrix - this allows fast term
evaluation and assembling but its drawback is a very high memory consumption
for high polynomial orders. Also, the polynomial order has to be uniform over
the whole (sub)domain where a field is defined. Related to polynomial orders,
tables with quadrature points for numerical integration are available or can be
generated as needed.

We are now working on implementing Nédélec and Raviart-Thomas vector bases in
order to support other kinds of PDEs, such as the Maxwell equations of
electromagnetism.

Once the equations are assembled, a number solvers can be used to solve the
problem. SfePy provides a unified interface to many standard codes, for example
UMFPACK \cite{R19}, PETSc \cite{R20}, Pysparse \cite{R21} as well as the solvers available
in SciPy. Various solver classes are supported: linear, nonlinear, eigenvalue,
optimization, and time stepping.

Besides the external solvers mentioned above, several solvers are implemented
directly in SfePy.

Nonlinear/optimization solvers:%
\begin{itemize}

\item 

The Newton solver with a backtracking line-search is the default solver for
all problems. For simplicity we use a unified approach to solve both the
linear and non-linear problems - former (should) converge in a single
nonlinear solver iteration.
\item 

The steepest descent optimization solver with a backtracking line-search can
be used as a fallback optimization solver when more sophisticated solvers
fail.
\end{itemize}

Time-stepping-related solvers:%
\begin{itemize}

\item 

The stationary solver is used to solve time-independent problems.
\item 

The equation sequence solver is a stationary solver that analyzes the
dependencies among equations and solves smaller blocks first. An example
would be the thermoelasticity problem described below, where the elasticity
equation depends on the load given by temperature distribution, but the
temperature distribution (Poisson equation) does not depend on
deformation. Then the temperature distribution can be found first, followed
by the elasticity problem with the already known temperature load. This
greatly reduces memory usage and improves speed of solution.
\item 

The simple implicit time stepping solver is used for (quasistatic)
time-dependent problems, using a fixed time step.
\item 

The adaptive implicit time stepping solver can change the time step according
to a user provided function. The default function \texttt{adapt\_time\_step()}
decreases the step in case of bad Newton convergence and increases the step
(up to a limit) when the convergence is fast. It is convenient for large
deformation (hyperelasticity) problems.
\item 

The explicit time stepping solver can be used for dynamic problems.
\end{itemize}

\section{Thermoelasticity Example%
  \label{thermoelasticity-example}%
}

This example involves calculating a \textbf{temperature distribution} in an object
followed by an \textbf{elastic deformation analysis} of the object loaded by the
thermal expansion and boundary displacement constraints. It shows how to use
SfePy in a script/interactively. The actual equations (weak form) are described
below. The entire script consists of the following steps:

Import modules. The SfePy package is organized into several sub-packages. The
example uses:%
\begin{itemize}

\item 

\texttt{sfepy.fem}: the finite element method (FEM) modules
\item 

\texttt{sfepy.terms}: the weak formulation terms - equations building
blocks
\item 

\texttt{sfepy.solvers}: interfaces to various solvers (SciPy, PETSc, ...)
\item 

\texttt{sfepy.postprocess}: post-processing \& visualization based on
Mayavi
\end{itemize}
\begin{Verbatim}[commandchars=\\\{\},fontsize=\footnotesize]
\PY{k+kn}{import} \PY{n+nn}{numpy} \PY{k+kn}{as} \PY{n+nn}{np}

\PY{k+kn}{from} \PY{n+nn}{sfepy.fem} \PY{k+kn}{import} \PY{p}{(}\PY{n}{Mesh}\PY{p}{,} \PY{n}{Domain}\PY{p}{,} \PY{n}{Field}\PY{p}{,}
                       \PY{n}{FieldVariable}\PY{p}{,}
                       \PY{n}{Material}\PY{p}{,} \PY{n}{Integral}\PY{p}{,}
                       \PY{n}{Equation}\PY{p}{,} \PY{n}{Equations}\PY{p}{,}
                       \PY{n}{ProblemDefinition}\PY{p}{)}
\PY{k+kn}{from} \PY{n+nn}{sfepy.terms} \PY{k+kn}{import} \PY{n}{Term}
\PY{k+kn}{from} \PY{n+nn}{sfepy.fem.conditions} \PY{k+kn}{import} \PY{n}{Conditions}\PY{p}{,} \PY{n}{EssentialBC}
\PY{k+kn}{from} \PY{n+nn}{sfepy.solvers.ls} \PY{k+kn}{import} \PY{n}{ScipyDirect}
\PY{k+kn}{from} \PY{n+nn}{sfepy.solvers.nls} \PY{k+kn}{import} \PY{n}{Newton}
\PY{k+kn}{from} \PY{n+nn}{sfepy.postprocess} \PY{k+kn}{import} \PY{n}{Viewer}
\end{Verbatim}
Load a mesh file defining the object geometry.\begin{Verbatim}[commandchars=\\\{\},fontsize=\footnotesize]
\PY{n}{mesh} \PY{o}{=} \PY{n}{Mesh}\PY{o}{.}\PY{n}{from\PYZus{}file}\PY{p}{(}\PY{l+s}{\PYZsq{}}\PY{l+s}{meshes/2d/square\PYZus{}tri2.mesh}\PY{l+s}{\PYZsq{}}\PY{p}{)}
\PY{n}{domain} \PY{o}{=} \PY{n}{Domain}\PY{p}{(}\PY{l+s}{\PYZsq{}}\PY{l+s}{domain}\PY{l+s}{\PYZsq{}}\PY{p}{,} \PY{n}{mesh}\PY{p}{)}
\end{Verbatim}
Define solution and boundary conditions domains, called regions.\begin{Verbatim}[commandchars=\\\{\},fontsize=\footnotesize]
\PY{n}{omega} \PY{o}{=} \PY{n}{domain}\PY{o}{.}\PY{n}{create\PYZus{}region}\PY{p}{(}\PY{l+s}{\PYZsq{}}\PY{l+s}{Omega}\PY{l+s}{\PYZsq{}}\PY{p}{,} \PY{l+s}{\PYZsq{}}\PY{l+s}{all}\PY{l+s}{\PYZsq{}}\PY{p}{)}
\PY{n}{left} \PY{o}{=} \PY{n}{domain}\PY{o}{.}\PY{n}{create\PYZus{}region}\PY{p}{(}\PY{l+s}{\PYZsq{}}\PY{l+s}{Left}\PY{l+s}{\PYZsq{}}\PY{p}{,}
                            \PY{l+s}{\PYZsq{}}\PY{l+s}{vertices in x \PYZlt{} \PYZhy{}0.999}\PY{l+s}{\PYZsq{}}\PY{p}{,}
                            \PY{l+s}{\PYZsq{}}\PY{l+s}{facet}\PY{l+s}{\PYZsq{}}\PY{p}{)}
\PY{n}{right} \PY{o}{=} \PY{n}{domain}\PY{o}{.}\PY{n}{create\PYZus{}region}\PY{p}{(}\PY{l+s}{\PYZsq{}}\PY{l+s}{Right}\PY{l+s}{\PYZsq{}}\PY{p}{,}
                             \PY{l+s}{\PYZsq{}}\PY{l+s}{vertices in x \PYZgt{} 0.999}\PY{l+s}{\PYZsq{}}\PY{p}{,}
                             \PY{l+s}{\PYZsq{}}\PY{l+s}{facet}\PY{l+s}{\PYZsq{}}\PY{p}{)}
\PY{n}{bottom} \PY{o}{=} \PY{n}{domain}\PY{o}{.}\PY{n}{create\PYZus{}region}\PY{p}{(}\PY{l+s}{\PYZsq{}}\PY{l+s}{Bottom}\PY{l+s}{\PYZsq{}}\PY{p}{,}
                              \PY{l+s}{\PYZsq{}}\PY{l+s}{vertices in y \PYZlt{} \PYZhy{}0.999}\PY{l+s}{\PYZsq{}}\PY{p}{,}
                              \PY{l+s}{\PYZsq{}}\PY{l+s}{facet}\PY{l+s}{\PYZsq{}}\PY{p}{)}
\PY{n}{top} \PY{o}{=} \PY{n}{domain}\PY{o}{.}\PY{n}{create\PYZus{}region}\PY{p}{(}\PY{l+s}{\PYZsq{}}\PY{l+s}{Top}\PY{l+s}{\PYZsq{}}\PY{p}{,}
                           \PY{l+s}{\PYZsq{}}\PY{l+s}{vertices in y \PYZgt{} 0.999}\PY{l+s}{\PYZsq{}}\PY{p}{,}
                           \PY{l+s}{\PYZsq{}}\PY{l+s}{facet}\PY{l+s}{\PYZsq{}}\PY{p}{)}
\end{Verbatim}
Save regions for visualization.\begin{Verbatim}[commandchars=\\\{\},fontsize=\footnotesize]
\PY{n}{domain}\PY{o}{.}\PY{n}{save\PYZus{}regions\PYZus{}as\PYZus{}groups}\PY{p}{(}\PY{l+s}{\PYZsq{}}\PY{l+s}{regions.vtk}\PY{l+s}{\PYZsq{}}\PY{p}{)}
\end{Verbatim}
Use a quadratic approximation for temperature field, define unknown $T$
and test $s$ variables.\begin{Verbatim}[commandchars=\\\{\},fontsize=\footnotesize]
\PY{n}{field\PYZus{}t} \PY{o}{=} \PY{n}{Field}\PY{o}{.}\PY{n}{from\PYZus{}args}\PY{p}{(}\PY{l+s}{\PYZsq{}}\PY{l+s}{temperature}\PY{l+s}{\PYZsq{}}\PY{p}{,} \PY{n}{np}\PY{o}{.}\PY{n}{float64}\PY{p}{,}
                          \PY{l+s}{\PYZsq{}}\PY{l+s}{scalar}\PY{l+s}{\PYZsq{}}\PY{p}{,} \PY{n}{omega}\PY{p}{,} \PY{l+m+mi}{2}\PY{p}{)}
\PY{n}{t} \PY{o}{=} \PY{n}{FieldVariable}\PY{p}{(}\PY{l+s}{\PYZsq{}}\PY{l+s}{t}\PY{l+s}{\PYZsq{}}\PY{p}{,} \PY{l+s}{\PYZsq{}}\PY{l+s}{unknown}\PY{l+s}{\PYZsq{}}\PY{p}{,} \PY{n}{field\PYZus{}t}\PY{p}{,} \PY{l+m+mi}{1}\PY{p}{)}
\PY{n}{s} \PY{o}{=} \PY{n}{FieldVariable}\PY{p}{(}\PY{l+s}{\PYZsq{}}\PY{l+s}{s}\PY{l+s}{\PYZsq{}}\PY{p}{,} \PY{l+s}{\PYZsq{}}\PY{l+s}{test}\PY{l+s}{\PYZsq{}}\PY{p}{,} \PY{n}{field\PYZus{}t}\PY{p}{,} \PY{l+m+mi}{1}\PY{p}{,}
                  \PY{n}{primary\PYZus{}var\PYZus{}name}\PY{o}{=}\PY{l+s}{\PYZsq{}}\PY{l+s}{t}\PY{l+s}{\PYZsq{}}\PY{p}{)}
\end{Verbatim}
Define numerical quadrature for the approximate integration rule.\begin{Verbatim}[commandchars=\\\{\},fontsize=\footnotesize]
\PY{n}{integral} \PY{o}{=} \PY{n}{Integral}\PY{p}{(}\PY{l+s}{\PYZsq{}}\PY{l+s}{i}\PY{l+s}{\PYZsq{}}\PY{p}{,} \PY{n}{order}\PY{o}{=}\PY{l+m+mi}{2}\PY{p}{)}
\end{Verbatim}
Define the Laplace equation governing the temperature distribution:\begin{equation*}
\int_{\Omega} \nabla s \cdot \nabla T = 0 \;, \quad \forall s \;.
\end{equation*}\begin{Verbatim}[commandchars=\\\{\},fontsize=\footnotesize]
\PY{n}{term} \PY{o}{=} \PY{n}{Term}\PY{o}{.}\PY{n}{new}\PY{p}{(}\PY{l+s}{\PYZsq{}}\PY{l+s}{dw\PYZus{}laplace(s, t)}\PY{l+s}{\PYZsq{}}\PY{p}{,} \PY{n}{integral}\PY{p}{,} \PY{n}{omega}\PY{p}{,}
                \PY{n}{s}\PY{o}{=}\PY{n}{s}\PY{p}{,} \PY{n}{t}\PY{o}{=}\PY{n}{t}\PY{p}{)}
\PY{n}{eq} \PY{o}{=} \PY{n}{Equation}\PY{p}{(}\PY{l+s}{\PYZsq{}}\PY{l+s}{temperature}\PY{l+s}{\PYZsq{}}\PY{p}{,} \PY{n}{term}\PY{p}{)}
\PY{n}{eqs} \PY{o}{=} \PY{n}{Equations}\PY{p}{(}\PY{p}{[}\PY{n}{eq}\PY{p}{]}\PY{p}{)}
\end{Verbatim}
Set boundary conditions for the temperature: $T = 10 \mbox{ on }
\Gamma_{\rm left}$, $T = 30 \mbox{ on } \Gamma_{\rm right}$.\begin{Verbatim}[commandchars=\\\{\},fontsize=\footnotesize]
\PY{n}{t\PYZus{}left} \PY{o}{=} \PY{n}{EssentialBC}\PY{p}{(}\PY{l+s}{\PYZsq{}}\PY{l+s}{t\PYZus{}left}\PY{l+s}{\PYZsq{}}\PY{p}{,}
                     \PY{n}{left}\PY{p}{,} \PY{p}{\PYZob{}}\PY{l+s}{\PYZsq{}}\PY{l+s}{t.0}\PY{l+s}{\PYZsq{}} \PY{p}{:} \PY{l+m+mf}{10.0}\PY{p}{\PYZcb{}}\PY{p}{)}
\PY{n}{t\PYZus{}right} \PY{o}{=} \PY{n}{EssentialBC}\PY{p}{(}\PY{l+s}{\PYZsq{}}\PY{l+s}{t\PYZus{}right}\PY{l+s}{\PYZsq{}}\PY{p}{,}
                      \PY{n}{right}\PY{p}{,} \PY{p}{\PYZob{}}\PY{l+s}{\PYZsq{}}\PY{l+s}{t.0}\PY{l+s}{\PYZsq{}} \PY{p}{:} \PY{l+m+mf}{30.0}\PY{p}{\PYZcb{}}\PY{p}{)}
\end{Verbatim}
Create linear (ScipyDirect) and nonlinear solvers (Newton).\begin{Verbatim}[commandchars=\\\{\},fontsize=\footnotesize]
\PY{n}{ls} \PY{o}{=} \PY{n}{ScipyDirect}\PY{p}{(}\PY{p}{\PYZob{}}\PY{p}{\PYZcb{}}\PY{p}{)}
\PY{n}{nls} \PY{o}{=} \PY{n}{Newton}\PY{p}{(}\PY{p}{\PYZob{}}\PY{p}{\PYZcb{}}\PY{p}{,} \PY{n}{lin\PYZus{}solver}\PY{o}{=}\PY{n}{ls}\PY{p}{)}
\end{Verbatim}
Combine the equations, boundary conditions and solvers to form a full problem
definition.\begin{Verbatim}[commandchars=\\\{\},fontsize=\footnotesize]
\PY{n}{pb} \PY{o}{=} \PY{n}{ProblemDefinition}\PY{p}{(}\PY{l+s}{\PYZsq{}}\PY{l+s}{temperature}\PY{l+s}{\PYZsq{}}\PY{p}{,} \PY{n}{equations}\PY{o}{=}\PY{n}{eqs}\PY{p}{,}
                       \PY{n}{nls}\PY{o}{=}\PY{n}{nls}\PY{p}{,} \PY{n}{ls}\PY{o}{=}\PY{n}{ls}\PY{p}{)}
\PY{n}{pb}\PY{o}{.}\PY{n}{time\PYZus{}update}\PY{p}{(}\PY{n}{ebcs}\PY{o}{=}\PY{n}{Conditions}\PY{p}{(}\PY{p}{[}\PY{n}{t\PYZus{}left}\PY{p}{,} \PY{n}{t\PYZus{}right}\PY{p}{]}\PY{p}{)}\PY{p}{)}
\end{Verbatim}
Solve the temperature distribution problem to get $T$.\begin{Verbatim}[commandchars=\\\{\},fontsize=\footnotesize]
\PY{n}{temperature} \PY{o}{=} \PY{n}{pb}\PY{o}{.}\PY{n}{solve}\PY{p}{(}\PY{p}{)}
\PY{n}{out} \PY{o}{=} \PY{n}{temperature}\PY{o}{.}\PY{n}{create\PYZus{}output\PYZus{}dict}\PY{p}{(}\PY{p}{)}
\end{Verbatim}
Use a linear approximation for displacement field, define unknown
$\underline{u}$ and test $\underline{v}$ variables. The variables
are vectors with two components in any point, as we are solving on a 2D domain.\begin{Verbatim}[commandchars=\\\{\},fontsize=\footnotesize]
\PY{n}{field\PYZus{}u} \PY{o}{=} \PY{n}{Field}\PY{o}{.}\PY{n}{from\PYZus{}args}\PY{p}{(}\PY{l+s}{\PYZsq{}}\PY{l+s}{displacement}\PY{l+s}{\PYZsq{}}\PY{p}{,} \PY{n}{np}\PY{o}{.}\PY{n}{float64}\PY{p}{,}
                          \PY{l+s}{\PYZsq{}}\PY{l+s}{vector}\PY{l+s}{\PYZsq{}}\PY{p}{,} \PY{n}{omega}\PY{p}{,} \PY{l+m+mi}{1}\PY{p}{)}
\PY{n}{u} \PY{o}{=} \PY{n}{FieldVariable}\PY{p}{(}\PY{l+s}{\PYZsq{}}\PY{l+s}{u}\PY{l+s}{\PYZsq{}}\PY{p}{,} \PY{l+s}{\PYZsq{}}\PY{l+s}{unknown}\PY{l+s}{\PYZsq{}}\PY{p}{,} \PY{n}{field\PYZus{}u}\PY{p}{,} \PY{n}{mesh}\PY{o}{.}\PY{n}{dim}\PY{p}{)}
\PY{n}{v} \PY{o}{=} \PY{n}{FieldVariable}\PY{p}{(}\PY{l+s}{\PYZsq{}}\PY{l+s}{v}\PY{l+s}{\PYZsq{}}\PY{p}{,} \PY{l+s}{\PYZsq{}}\PY{l+s}{test}\PY{l+s}{\PYZsq{}}\PY{p}{,} \PY{n}{field\PYZus{}u}\PY{p}{,} \PY{n}{mesh}\PY{o}{.}\PY{n}{dim}\PY{p}{,}
                  \PY{n}{primary\PYZus{}var\PYZus{}name}\PY{o}{=}\PY{l+s}{\PYZsq{}}\PY{l+s}{u}\PY{l+s}{\PYZsq{}}\PY{p}{)}
\end{Verbatim}
Set Lamé parameters of elasticity $\lambda$, $\mu$, thermal
expansion coefficient $\alpha_{ij}$ and background temperature
$T_0$. Constant values are used here. In general, material parameters can
be given as functions of space and time.\begin{Verbatim}[commandchars=\\\{\},fontsize=\footnotesize]
\PY{n}{lam} \PY{o}{=} \PY{l+m+mf}{10.0} \PY{c}{\PYZsh{} Lame parameters.}
\PY{n}{mu} \PY{o}{=} \PY{l+m+mf}{5.0}
\PY{n}{te} \PY{o}{=} \PY{l+m+mf}{0.5} \PY{c}{\PYZsh{} Thermal expansion coefficient.}
\PY{n}{T0} \PY{o}{=} \PY{l+m+mf}{20.0} \PY{c}{\PYZsh{} Background temperature.}
\PY{n}{eye\PYZus{}sym} \PY{o}{=} \PY{n}{np}\PY{o}{.}\PY{n}{array}\PY{p}{(}\PY{p}{[}\PY{p}{[}\PY{l+m+mi}{1}\PY{p}{]}\PY{p}{,} \PY{p}{[}\PY{l+m+mi}{1}\PY{p}{]}\PY{p}{,} \PY{p}{[}\PY{l+m+mi}{0}\PY{p}{]}\PY{p}{]}\PY{p}{,}
                   \PY{n}{dtype}\PY{o}{=}\PY{n}{np}\PY{o}{.}\PY{n}{float64}\PY{p}{)}
\PY{n}{m} \PY{o}{=} \PY{n}{Material}\PY{p}{(}\PY{l+s}{\PYZsq{}}\PY{l+s}{m}\PY{l+s}{\PYZsq{}}\PY{p}{,} \PY{n}{lam}\PY{o}{=}\PY{n}{lam}\PY{p}{,} \PY{n}{mu}\PY{o}{=}\PY{n}{mu}\PY{p}{,}
             \PY{n}{alpha}\PY{o}{=}\PY{n}{te} \PY{o}{*} \PY{n}{eye\PYZus{}sym}\PY{p}{)}
\end{Verbatim}
Define and set the temperature load variable to $T - T_0$.\begin{Verbatim}[commandchars=\\\{\},fontsize=\footnotesize]
\PY{n}{t2} \PY{o}{=} \PY{n}{FieldVariable}\PY{p}{(}\PY{l+s}{\PYZsq{}}\PY{l+s}{t}\PY{l+s}{\PYZsq{}}\PY{p}{,} \PY{l+s}{\PYZsq{}}\PY{l+s}{parameter}\PY{l+s}{\PYZsq{}}\PY{p}{,} \PY{n}{field\PYZus{}t}\PY{p}{,} \PY{l+m+mi}{1}\PY{p}{,}
                   \PY{n}{primary\PYZus{}var\PYZus{}name}\PY{o}{=}\PY{l+s}{\PYZsq{}}\PY{l+s}{(set\PYZhy{}to\PYZhy{}None)}\PY{l+s}{\PYZsq{}}\PY{p}{)}
\PY{n}{t2}\PY{o}{.}\PY{n}{set\PYZus{}data}\PY{p}{(}\PY{n}{t}\PY{p}{(}\PY{p}{)} \PY{o}{\PYZhy{}} \PY{n}{T0}\PY{p}{)}
\end{Verbatim}
Define the thermoelasticity equation governing structure deformation:\begin{equation*}
\int_{\Omega} D_{ijkl}\ e_{ij}(\underline{v}) e_{kl}(\underline{u}) -
\int_{\Omega} (T - T_0)\ \alpha_{ij} e_{ij}(\underline{v}) = 0 \;, \quad
\forall \underline{v} \;,
\end{equation*}where $D_{ijkl} = \mu (\delta_{ik} \delta_{jl}+\delta_{il} \delta_{jk}) +
\lambda \ \delta_{ij} \delta_{kl}$ is the homogeneous isotropic elasticity
tensor and $e_{ij}(\underline{u}) = \frac{1}{2}(\frac{\partial
u_i}{\partial x_j} + \frac{\partial u_j}{\partial x_i})$ is the small strain
tensor. The equations can be built as linear combinations of terms.\begin{Verbatim}[commandchars=\\\{\},fontsize=\footnotesize]
\PY{n}{term1} \PY{o}{=} \PY{n}{Term}\PY{o}{.}\PY{n}{new}\PY{p}{(}\PY{l+s}{\PYZsq{}}\PY{l+s}{dw\PYZus{}lin\PYZus{}elastic\PYZus{}iso(m.lam, m.mu, v, u)}\PY{l+s}{\PYZsq{}}\PY{p}{,}
                 \PY{n}{integral}\PY{p}{,} \PY{n}{omega}\PY{p}{,} \PY{n}{m}\PY{o}{=}\PY{n}{m}\PY{p}{,} \PY{n}{v}\PY{o}{=}\PY{n}{v}\PY{p}{,} \PY{n}{u}\PY{o}{=}\PY{n}{u}\PY{p}{)}
\PY{n}{term2} \PY{o}{=} \PY{n}{Term}\PY{o}{.}\PY{n}{new}\PY{p}{(}\PY{l+s}{\PYZsq{}}\PY{l+s}{dw\PYZus{}biot(m.alpha, v, t)}\PY{l+s}{\PYZsq{}}\PY{p}{,}
                 \PY{n}{integral}\PY{p}{,} \PY{n}{omega}\PY{p}{,} \PY{n}{m}\PY{o}{=}\PY{n}{m}\PY{p}{,} \PY{n}{v}\PY{o}{=}\PY{n}{v}\PY{p}{,} \PY{n}{t}\PY{o}{=}\PY{n}{t2}\PY{p}{)}
\PY{n}{eq} \PY{o}{=} \PY{n}{Equation}\PY{p}{(}\PY{l+s}{\PYZsq{}}\PY{l+s}{temperature}\PY{l+s}{\PYZsq{}}\PY{p}{,} \PY{n}{term1} \PY{o}{\PYZhy{}} \PY{n}{term2}\PY{p}{)}
\PY{n}{eqs} \PY{o}{=} \PY{n}{Equations}\PY{p}{(}\PY{p}{[}\PY{n}{eq}\PY{p}{]}\PY{p}{)}
\end{Verbatim}
Set boundary conditions for the displacements: $\underline{u} = 0 \mbox{
on } \Gamma_{\rm bottom}$, $u_1 = 0.0 \mbox{ on } \Gamma_{\rm top}$
($x$ -component).\begin{Verbatim}[commandchars=\\\{\},fontsize=\footnotesize]
\PY{n}{u\PYZus{}bottom} \PY{o}{=} \PY{n}{EssentialBC}\PY{p}{(}\PY{l+s}{\PYZsq{}}\PY{l+s}{u\PYZus{}bottom}\PY{l+s}{\PYZsq{}}\PY{p}{,}
                       \PY{n}{bottom}\PY{p}{,} \PY{p}{\PYZob{}}\PY{l+s}{\PYZsq{}}\PY{l+s}{u.all}\PY{l+s}{\PYZsq{}} \PY{p}{:} \PY{l+m+mf}{0.0}\PY{p}{\PYZcb{}}\PY{p}{)}
\PY{n}{u\PYZus{}top} \PY{o}{=} \PY{n}{EssentialBC}\PY{p}{(}\PY{l+s}{\PYZsq{}}\PY{l+s}{u\PYZus{}top}\PY{l+s}{\PYZsq{}}\PY{p}{,}
                    \PY{n}{top}\PY{p}{,} \PY{p}{\PYZob{}}\PY{l+s}{\PYZsq{}}\PY{l+s}{u.[0]}\PY{l+s}{\PYZsq{}} \PY{p}{:} \PY{l+m+mf}{0.0}\PY{p}{\PYZcb{}}\PY{p}{)}
\end{Verbatim}
Set the thermoelasticity equations and boundary conditions to the problem
definition.\begin{Verbatim}[commandchars=\\\{\},fontsize=\footnotesize]
\PY{n}{pb}\PY{o}{.}\PY{n}{set\PYZus{}equations\PYZus{}instance}\PY{p}{(}\PY{n}{eqs}\PY{p}{,} \PY{n}{keep\PYZus{}solvers}\PY{o}{=}\PY{n+nb+bp}{True}\PY{p}{)}
\PY{n}{pb}\PY{o}{.}\PY{n}{time\PYZus{}update}\PY{p}{(}\PY{n}{ebcs}\PY{o}{=}\PY{n}{Conditions}\PY{p}{(}\PY{p}{[}\PY{n}{u\PYZus{}bottom}\PY{p}{,} \PY{n}{u\PYZus{}top}\PY{p}{]}\PY{p}{)}\PY{p}{)}
\end{Verbatim}
Solve the thermoelasticity problem to get $\underline{u}$.\begin{Verbatim}[commandchars=\\\{\},fontsize=\footnotesize]
\PY{n}{displacement} \PY{o}{=} \PY{n}{pb}\PY{o}{.}\PY{n}{solve}\PY{p}{(}\PY{p}{)}
\PY{n}{out}\PY{o}{.}\PY{n}{update}\PY{p}{(}\PY{n}{displacement}\PY{o}{.}\PY{n}{create\PYZus{}output\PYZus{}dict}\PY{p}{(}\PY{p}{)}\PY{p}{)}
\end{Verbatim}
Save the solution of both problems into a single VTK file.\begin{Verbatim}[commandchars=\\\{\},fontsize=\footnotesize]
\PY{n}{pb}\PY{o}{.}\PY{n}{save\PYZus{}state}\PY{p}{(}\PY{l+s}{\PYZsq{}}\PY{l+s}{thermoelasticity.vtk}\PY{l+s}{\PYZsq{}}\PY{p}{,} \PY{n}{out}\PY{o}{=}\PY{n}{out}\PY{p}{)}
\end{Verbatim}
Display the solution using Mayavi.\begin{Verbatim}[commandchars=\\\{\},fontsize=\footnotesize]
\PY{n}{view} \PY{o}{=} \PY{n}{Viewer}\PY{p}{(}\PY{l+s}{\PYZsq{}}\PY{l+s}{thermoelasticity.vtk}\PY{l+s}{\PYZsq{}}\PY{p}{)}
\PY{n}{view}\PY{p}{(}\PY{n}{vector\PYZus{}mode}\PY{o}{=}\PY{l+s}{\PYZsq{}}\PY{l+s}{warp\PYZus{}norm}\PY{l+s}{\PYZsq{}}\PY{p}{,}
     \PY{n}{rel\PYZus{}scaling}\PY{o}{=}\PY{l+m+mi}{1}\PY{p}{,} \PY{n}{is\PYZus{}scalar\PYZus{}bar}\PY{o}{=}\PY{n+nb+bp}{True}\PY{p}{,}
     \PY{n}{is\PYZus{}wireframe}\PY{o}{=}\PY{n+nb+bp}{True}\PY{p}{,}
     \PY{n}{opacity}\PY{o}{=}\PY{p}{\PYZob{}}\PY{l+s}{\PYZsq{}}\PY{l+s}{wireframe}\PY{l+s}{\PYZsq{}} \PY{p}{:} \PY{l+m+mf}{0.1}\PY{p}{\PYZcb{}}\PY{p}{)}
\end{Verbatim}

\subsection{Results%
  \label{results}%
}
The above script saves the domain geometry as well as the temperature and
displacement fields into a VTK file called \texttt{'thermoelasticity.vtk'} and also
displays the results using Mayavi. The results are shown in Figures
\DUrole{ref}{temperature} and \DUrole{ref}{displacement}.\begin{figure}[h]\noindent\makebox[\columnwidth][c]{\includegraphics[scale=0.20]{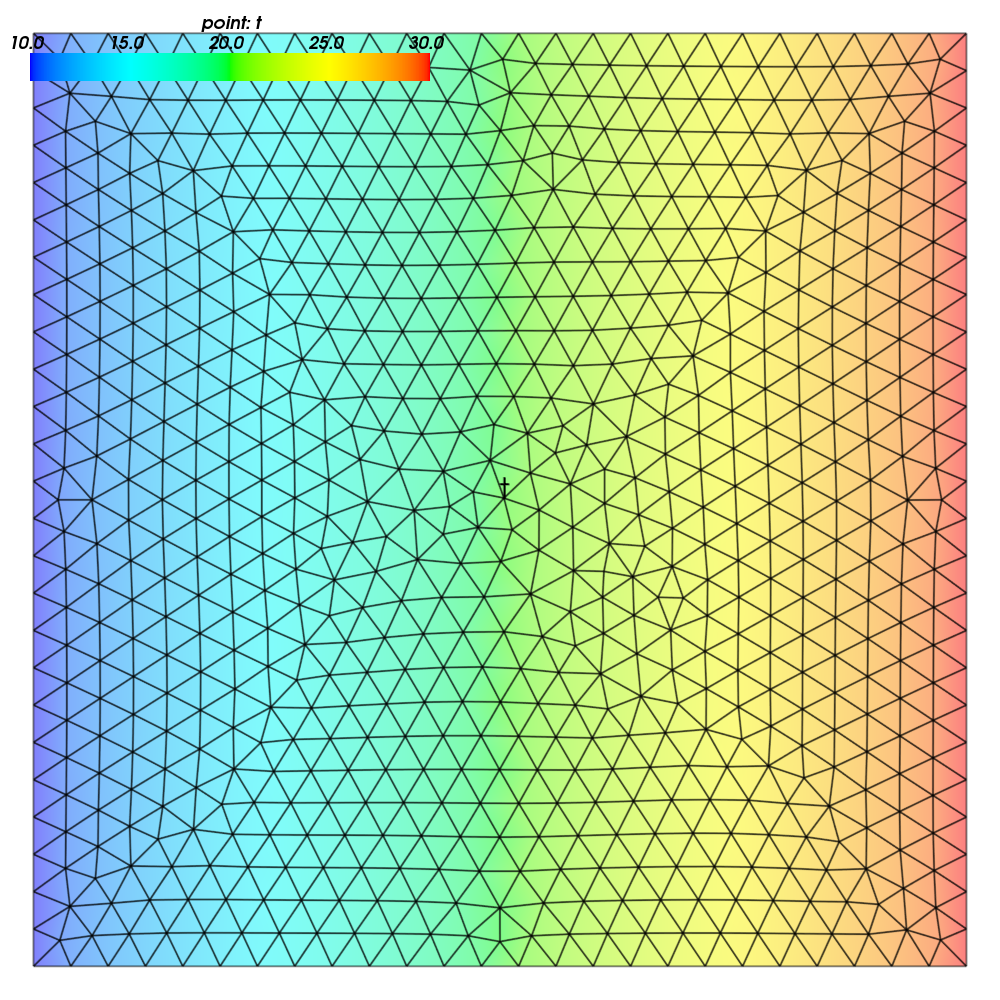}}
\caption{The temperature distribution. \DUrole{label}{temperature}}
\end{figure}\begin{figure}[h]\noindent\makebox[\columnwidth][c]{\includegraphics[scale=0.20]{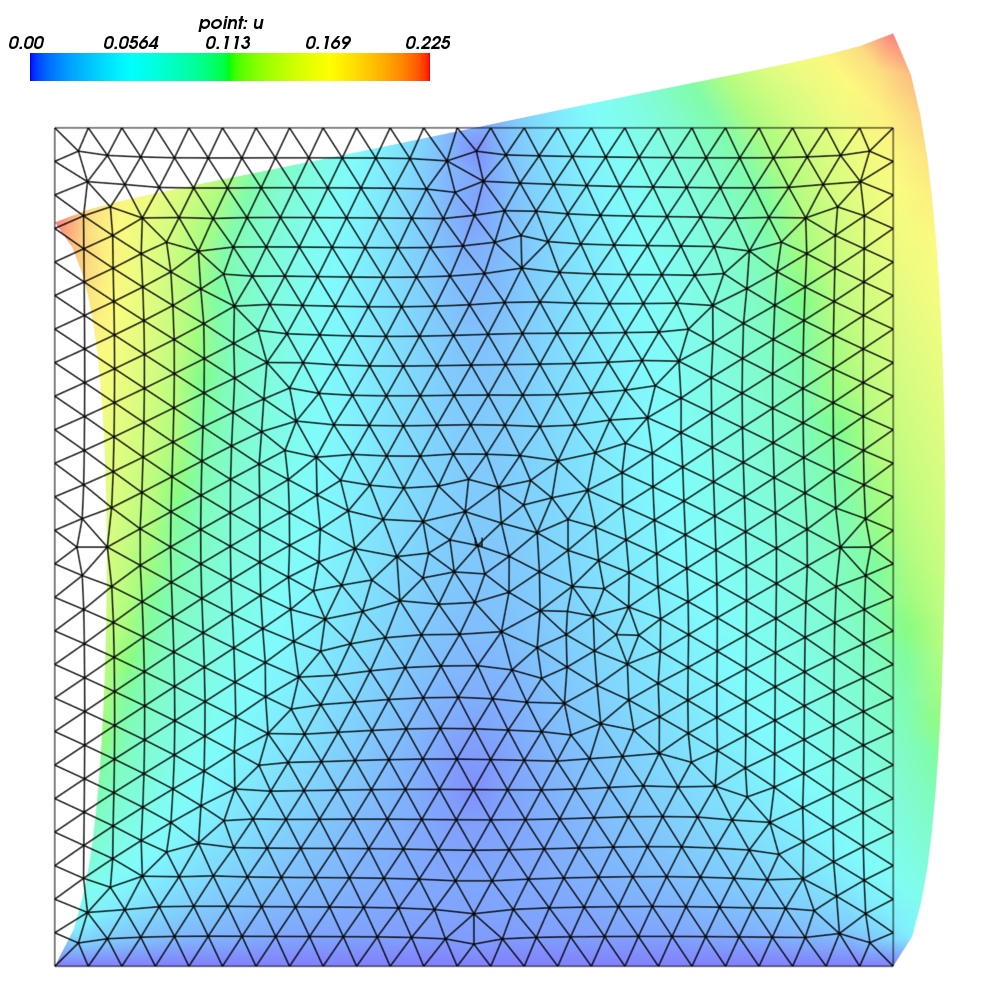}}
\caption{The deformed mesh showing displacements. \DUrole{label}{displacement}}
\end{figure}

\section{Alternative Way: Problem Description Files%
  \label{alternative-way-problem-description-files}%
}

Problem description files (PDF) are Python modules containing definitions of
the various components (mesh, regions, fields, equations, ...)  using basic
data types such as \texttt{dict} and \texttt{tuple}. For simple problems, no programming
at all is required. On the other hand, all the power of Python (and supporting
SfePy modules) is available when needed. The definitions are used to construct
and initialize in an automatic way the corresponding objects, similarly to what
was presented in the example above, and the problem is solved. The main script
for running a simulation described in a PDF is called \texttt{simple.py}.

\subsection{Example: Temperature Distribution%
  \label{example-temperature-distribution}%
}

This example defines the problem of temperature distribution on a 2D
rectangular domain. It directly corresponds to the temperature part of the
thermoelasticity example, only for the sake of completeness a definition of
a material coefficient is shown as well.\begin{Verbatim}[commandchars=\\\{\},fontsize=\footnotesize]
\PY{k+kn}{from} \PY{n+nn}{sfepy} \PY{k+kn}{import} \PY{n}{data\PYZus{}dir}
\PY{n}{filename\PYZus{}mesh} \PY{o}{=} \PY{n}{data\PYZus{}dir} \PY{o}{+} \PY{l+s}{\PYZsq{}}\PY{l+s}{/meshes/2d/square\PYZus{}tri2.mesh}\PY{l+s}{\PYZsq{}}

\PY{n}{materials} \PY{o}{=} \PY{p}{\PYZob{}}
    \PY{l+s}{\PYZsq{}}\PY{l+s}{coef}\PY{l+s}{\PYZsq{}} \PY{p}{:} \PY{p}{(}\PY{p}{\PYZob{}}\PY{l+s}{\PYZsq{}}\PY{l+s}{val}\PY{l+s}{\PYZsq{}} \PY{p}{:} \PY{l+m+mf}{1.0}\PY{p}{\PYZcb{}}\PY{p}{,}\PY{p}{)}\PY{p}{,}
\PY{p}{\PYZcb{}}

\PY{n}{regions} \PY{o}{=} \PY{p}{\PYZob{}}
    \PY{l+s}{\PYZsq{}}\PY{l+s}{Omega}\PY{l+s}{\PYZsq{}} \PY{p}{:} \PY{l+s}{\PYZsq{}}\PY{l+s}{all}\PY{l+s}{\PYZsq{}}\PY{p}{,}
    \PY{l+s}{\PYZsq{}}\PY{l+s}{Left}\PY{l+s}{\PYZsq{}} \PY{p}{:} \PY{p}{(}\PY{l+s}{\PYZsq{}}\PY{l+s}{vertices in (x \PYZlt{} \PYZhy{}0.999)}\PY{l+s}{\PYZsq{}}\PY{p}{,} \PY{l+s}{\PYZsq{}}\PY{l+s}{facet}\PY{l+s}{\PYZsq{}}\PY{p}{)}\PY{p}{,}
    \PY{l+s}{\PYZsq{}}\PY{l+s}{Right}\PY{l+s}{\PYZsq{}} \PY{p}{:} \PY{p}{(}\PY{l+s}{\PYZsq{}}\PY{l+s}{vertices in (x \PYZgt{} 0.999)}\PY{l+s}{\PYZsq{}}\PY{p}{,} \PY{l+s}{\PYZsq{}}\PY{l+s}{facet}\PY{l+s}{\PYZsq{}}\PY{p}{)}\PY{p}{,}
\PY{p}{\PYZcb{}}

\PY{n}{fields} \PY{o}{=} \PY{p}{\PYZob{}}
    \PY{l+s}{\PYZsq{}}\PY{l+s}{temperature}\PY{l+s}{\PYZsq{}} \PY{p}{:} \PY{p}{(}\PY{l+s}{\PYZsq{}}\PY{l+s}{real}\PY{l+s}{\PYZsq{}}\PY{p}{,} \PY{l+m+mi}{1}\PY{p}{,} \PY{l+s}{\PYZsq{}}\PY{l+s}{Omega}\PY{l+s}{\PYZsq{}}\PY{p}{,} \PY{l+m+mi}{2}\PY{p}{)}\PY{p}{,}
\PY{p}{\PYZcb{}}

\PY{n}{variables} \PY{o}{=} \PY{p}{\PYZob{}}
    \PY{l+s}{\PYZsq{}}\PY{l+s}{t}\PY{l+s}{\PYZsq{}} \PY{p}{:} \PY{p}{(}\PY{l+s}{\PYZsq{}}\PY{l+s}{unknown field}\PY{l+s}{\PYZsq{}}\PY{p}{,} \PY{l+s}{\PYZsq{}}\PY{l+s}{temperature}\PY{l+s}{\PYZsq{}}\PY{p}{,} \PY{l+m+mi}{0}\PY{p}{)}\PY{p}{,}
    \PY{l+s}{\PYZsq{}}\PY{l+s}{s}\PY{l+s}{\PYZsq{}} \PY{p}{:} \PY{p}{(}\PY{l+s}{\PYZsq{}}\PY{l+s}{test field}\PY{l+s}{\PYZsq{}}\PY{p}{,}    \PY{l+s}{\PYZsq{}}\PY{l+s}{temperature}\PY{l+s}{\PYZsq{}}\PY{p}{,} \PY{l+s}{\PYZsq{}}\PY{l+s}{t}\PY{l+s}{\PYZsq{}}\PY{p}{)}\PY{p}{,}
\PY{p}{\PYZcb{}}

\PY{n}{ebcs} \PY{o}{=} \PY{p}{\PYZob{}}
    \PY{l+s}{\PYZsq{}}\PY{l+s}{t\PYZus{}left}\PY{l+s}{\PYZsq{}} \PY{p}{:} \PY{p}{(}\PY{l+s}{\PYZsq{}}\PY{l+s}{Left}\PY{l+s}{\PYZsq{}}\PY{p}{,} \PY{p}{\PYZob{}}\PY{l+s}{\PYZsq{}}\PY{l+s}{t.0}\PY{l+s}{\PYZsq{}} \PY{p}{:} \PY{l+m+mf}{10.0}\PY{p}{\PYZcb{}}\PY{p}{)}\PY{p}{,}
    \PY{l+s}{\PYZsq{}}\PY{l+s}{t\PYZus{}right}\PY{l+s}{\PYZsq{}} \PY{p}{:} \PY{p}{(}\PY{l+s}{\PYZsq{}}\PY{l+s}{Right}\PY{l+s}{\PYZsq{}}\PY{p}{,} \PY{p}{\PYZob{}}\PY{l+s}{\PYZsq{}}\PY{l+s}{t.0}\PY{l+s}{\PYZsq{}} \PY{p}{:} \PY{l+m+mf}{30.0}\PY{p}{\PYZcb{}}\PY{p}{)}\PY{p}{,}
\PY{p}{\PYZcb{}}

\PY{n}{integrals} \PY{o}{=} \PY{p}{\PYZob{}}
    \PY{l+s}{\PYZsq{}}\PY{l+s}{i1}\PY{l+s}{\PYZsq{}} \PY{p}{:} \PY{p}{(}\PY{l+s}{\PYZsq{}}\PY{l+s}{v}\PY{l+s}{\PYZsq{}}\PY{p}{,} \PY{l+m+mi}{2}\PY{p}{)}\PY{p}{,}
\PY{p}{\PYZcb{}}

\PY{n}{equations} \PY{o}{=} \PY{p}{\PYZob{}}
    \PY{l+s}{\PYZsq{}}\PY{l+s}{eq}\PY{l+s}{\PYZsq{}} \PY{p}{:} \PY{l+s}{\PYZsq{}}\PY{l+s}{dw\PYZus{}laplace.i1.Omega(coef.val, s, t) = 0}\PY{l+s}{\PYZsq{}}
\PY{p}{\PYZcb{}}

\PY{n}{solvers} \PY{o}{=} \PY{p}{\PYZob{}}
    \PY{l+s}{\PYZsq{}}\PY{l+s}{ls}\PY{l+s}{\PYZsq{}} \PY{p}{:} \PY{p}{(}\PY{l+s}{\PYZsq{}}\PY{l+s}{ls.scipy\PYZus{}direct}\PY{l+s}{\PYZsq{}}\PY{p}{,} \PY{p}{\PYZob{}}\PY{p}{\PYZcb{}}\PY{p}{)}\PY{p}{,}
    \PY{l+s}{\PYZsq{}}\PY{l+s}{newton}\PY{l+s}{\PYZsq{}} \PY{p}{:} \PY{p}{(}\PY{l+s}{\PYZsq{}}\PY{l+s}{nls.newton}\PY{l+s}{\PYZsq{}}\PY{p}{,}
                \PY{p}{\PYZob{}}\PY{l+s}{\PYZsq{}}\PY{l+s}{i\PYZus{}max}\PY{l+s}{\PYZsq{}}      \PY{p}{:} \PY{l+m+mi}{1}\PY{p}{,}
                 \PY{l+s}{\PYZsq{}}\PY{l+s}{eps\PYZus{}a}\PY{l+s}{\PYZsq{}}      \PY{p}{:} \PY{l+m+mf}{1e\PYZhy{}10}\PY{p}{,}
    \PY{p}{\PYZcb{}}\PY{p}{)}\PY{p}{,}
\PY{p}{\PYZcb{}}

\PY{n}{options} \PY{o}{=} \PY{p}{\PYZob{}}
    \PY{l+s}{\PYZsq{}}\PY{l+s}{nls}\PY{l+s}{\PYZsq{}} \PY{p}{:} \PY{l+s}{\PYZsq{}}\PY{l+s}{newton}\PY{l+s}{\PYZsq{}}\PY{p}{,}
    \PY{l+s}{\PYZsq{}}\PY{l+s}{ls}\PY{l+s}{\PYZsq{}} \PY{p}{:} \PY{l+s}{\PYZsq{}}\PY{l+s}{ls}\PY{l+s}{\PYZsq{}}\PY{p}{,}
\PY{p}{\PYZcb{}}
\end{Verbatim}
Many more examples can be found at \url{http://docs.sfepy.org/gallery/gallery} or
\url{http://sfepy.org/doc-devel/examples.html}.\begin{figure*}[]\noindent\makebox[\textwidth][c]{\includegraphics[scale=0.90]{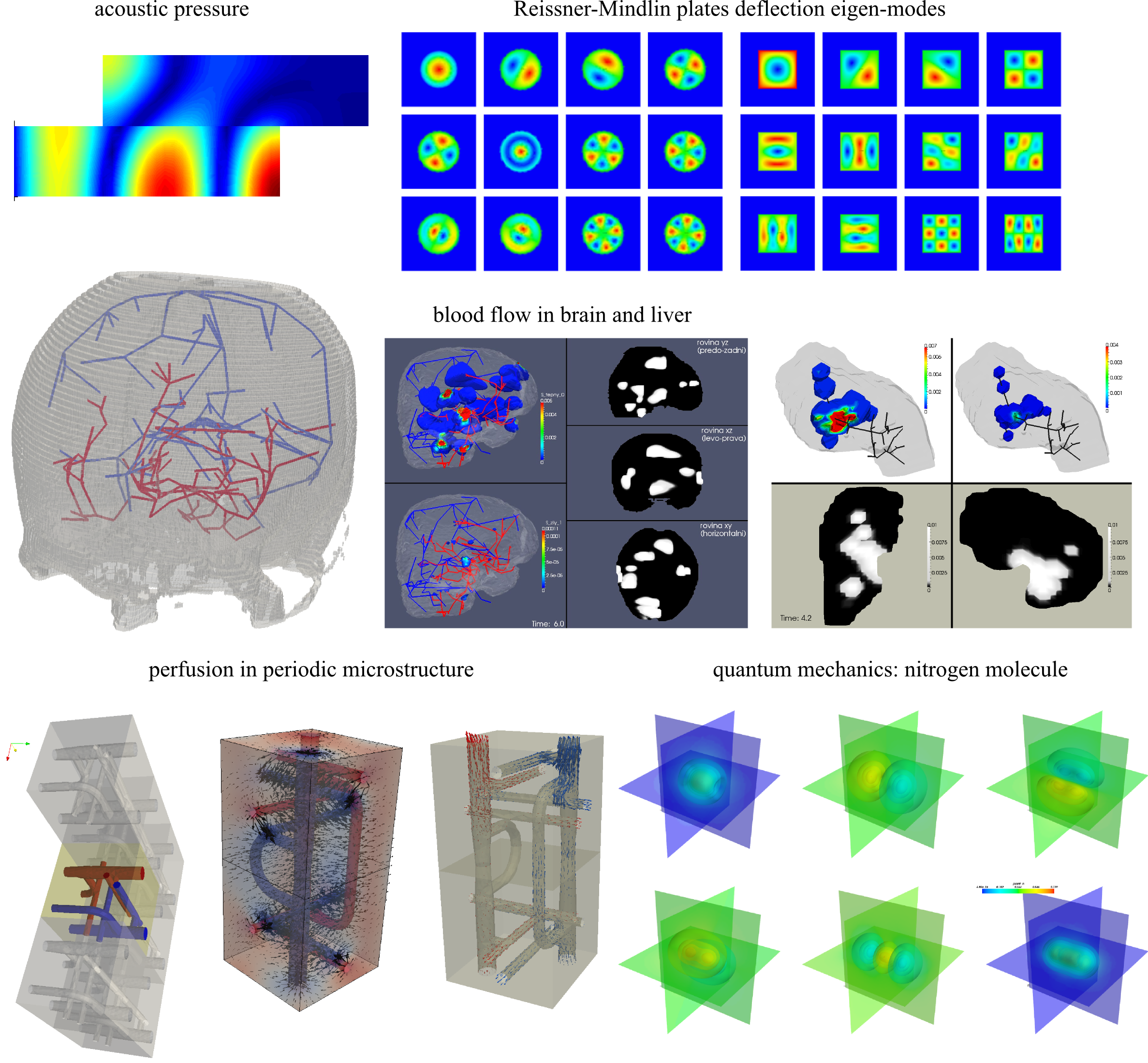}}
\caption{Gallery of applications. Perfusion and acoustic images by Vladimír
Lukeš. \DUrole{label}{gallery}}
\end{figure*}

\section{Conclusion%
  \label{conclusion}%
}

We briefly introduced the open source finite element package SfePy as a tool
for building domain-specific FE-based solvers as well as a black-box PDE
solver.

\subsection{Support%
  \label{support}%
}

Work on SfePy is partially supported by the Grant Agency of the Czech Republic,
projects P108/11/0853 and 101/09/1630.

\end{document}